%% file: nsml_main.tex
\documentclass{article}

\usepackage{microtype}
\usepackage{graphicx}
\usepackage{caption}
\usepackage{subcaption}
\usepackage{booktabs} 
\usepackage{xcolor}
\usepackage{kotex}
\usepackage[utf8]{inputenc}
\usepackage{listings}

\usepackage{hyperref}

\usepackage[accepted]{sysml2019}

\sysmltitlerunning{NSML: Meet the MLaaS platform with a real-world case study}

\begin{document}

\twocolumn[
\sysmltitle{NSML: Meet the MLaaS platform with a real-world case study}



\sysmlsetsymbol{equal}{*}

\begin{sysmlauthorlist}
\sysmlauthor{Hanjoo Kim}{equal,naver}
\sysmlauthor{Minkyu Kim}{equal,naver}
\sysmlauthor{Dongjoo Seo}{equal,naver}
\sysmlauthor{Jinwoong Kim}{naver}
\sysmlauthor{Heungseok Park}{naver}
\sysmlauthor{Soeun Park}{naver}
\sysmlauthor{Hyunwoo Jo}{naver}
\sysmlauthor{KyungHyun Kim}{naver}
\sysmlauthor{Youngil Yang}{naver}
\sysmlauthor{Youngkwan Kim}{naver}
\sysmlauthor{Nako Sung}{naver}
\sysmlauthor{Jung-Woo Ha}{naver}
\end{sysmlauthorlist}

\sysmlaffiliation{naver}{CLOVA AI Research, NAVER Corporation, Republic of Korea}

\sysmlcorrespondingauthor{Hanjoo Kim}{hanjoo.kim@navercorp.com}
\sysmlcorrespondingauthor{Jung-Woo Ha}{jungwoo.ha@navercorp.com}

\sysmlkeywords{Machine Learning, Cluster Computing, System, SysML}

\vskip 0.3in

\begin{abstract}
The boom of deep learning induced many industries and academies to introduce machine learning based approaches into their concern, competitively. However, existing machine learning frameworks are limited to sufficiently fulfill the collaboration and management for both data and models. We proposed NSML, a machine learning as a service (MLaaS) platform, to meet these demands. NSML helps machine learning work be easily launched on a NSML cluster and provides a collaborative environment which can afford development at enterprise scale. Finally, NSML users can deploy their own commercial services with NSML cluster. In addition, NSML furnishes convenient visualization tools which assist the users in analyzing their work. To verify the usefulness and accessibility of NSML, we performed some experiments with common examples. Furthermore, we examined the collaborative advantages of NSML through three competitions with real-world use cases.
\end{abstract}
]



\printAffiliationsAndNotice{\sysmlEqualContribution} 

\input{body.tex}

\bibliography{bibliography}
\bibliographystyle{sysml2019}
\end{document}

%% file: body.tex
\section{Introduction}
Starting with the success of convolutional neural network~\cite{deepimage}, deep learning has infiltrated into many fields such as image recognition, natural language processing, recommendation system, question-answering, and bio-medical applications.~\cite{yamada2018shakedrop,xiong2018microsoft,liang2018variational,yu2018qanet,deepbio}
As the applications of the deep learning grow, the complexity of the tasks for designing and tuning the neural network architectures increases as well as the size of models. Moreover, if we consider collaboration and computing resources for practical scale in real world, proper tools and platforms should be required to efficiently managing them. Many companies and organizations have constructed their own clusters and collaborative tools for developing machine learning models.~\cite{azure,googlecloud} There are still two main issues to build the total solution for machine learning tasks.

The first one is building the environment for training and validating the models. Most machine learning models have dependency on existing libraries and frameworks with various versions. For common user systems, these libraries conflict with each other known as ``dependency hell''.~\cite{Boettiger:2015:IDR:2723872.2723882} Consequently, the environmental setup is delayed or failed even before the main task begins. The other one is about collaborative work. It is essential that many users work on one huge machine learning task for enterprise-level projects. They may want to share the data and models, and reproduce colleagues' work with the same setup. If those components are not well managed, the task progress will not go smooth due to the insufficient and inefficient computing and storage resources.

Although some frameworks have been proposed to improve the efficiency of machine learning tasks, they are somewhat lacking in full-stack machine learning framework. For instance,   well known deep learning libraries such as TensorFlow~\cite{tensorflow} and PyTorch~\cite{pytorch} function as only deep learning model developing library and not collaborative tools nor cluster managers. On the other hand, Comet\footnote{\url{https://www.comet.ml/}}, Valohai\footnote{\url{https://valohai.com/}}, and FloydHub\footnote{\url{https://www.floydhub.com/}} provide management tools for machine learning work with well-designed user interface. Nevertheless, they rely on AWS(Amazon Web Service)~\cite{aws} or Google's cloud system as computing resources, and does not support the private computing cluster that users already possess.

In summary, we realize that the platform for machine learning solutions need to meet those requirements including 1) automatic computing resources allocation/release, 2) job scheduling and migration, 3) sharing or private options for source code and data, 4) monitoring function of keeping track of model learning status, 5) flexible hyperparameter tuning in training time, 6) comparison of the performances of many diverse hyperparameter models, and 7) a leaderboard for comparing models with others. As a result, we propose a brand-new machine learning platform, Naver Smart Machine Learning (NSML) for practical usages. Our platform is not only satisfying the users' needs, but also proved its usefulness and robustness with the real use cases. Our key contributions are summarized as follows:

\begin{itemize}
    \setlength\itemsep{0pt}
	\item{Efficient resource management}
	\item{Supporting for collaborative work}
    \item{Easy to deploy model}
   	\item{Case study with real-world examples}
    \item{Few code insertion for running on NSML}
	\item{An intuitive UI/UX for machine learning developers and researchers}
\end{itemize}

\section{Related Work}
\subsection{Platforms for machine learning tasks}
When many researchers start to implement deep neural networks, they often encounter trouble in setup for both hardware and software environments. The difficulty of building the environment may hurt the shareability and utilization of the computational resources, and beginners cannot even start the very first their own machine learning code. Many other machine learning platforms were proposed to mitigate this problem. These platforms helped users to focus on their main subjects, such as designing machine learning models. 

Most of the existing machine learning platforms share following features; reproducibility, usability, visualization, and model deployment. For instance, FloydHub uses Docker~\cite{docker} to package the machine learning job for sharing the configuration of machine learning model and its environment. With this feature, users can reproduce their machine learning (ML) tasks from any other environment without setting up manually. Meanwhile, Google Cloud Vision\footnote{\url{https://cloud.google.com/vision/}} supports RESTful API for deploying users' models, ensuring usability. Besides, it also supports Tensorboard~\cite{girija2016tensorflow} that visualizes both performance of models and model architecture itself. TensorFlow extended~\cite{baylor2017tfx} is also one of the most recent platform for machine learning from Google. Even if the most interesting part is the dataset management, however, its dependency on TensorFlow may limit availability.

\subsection{Cloud Platform}
Software-based services are increasing exponentially in many areas. As a result, the demand for computing resources for developing the services has increased, and a cloud platform has been created to facilitate the deployment. Cloud platform provides users scalability and flexibility, tailored to the users' taste, by providing virtual machines rather than physical machines.
Examples of such platforms include Amazon Web Service (AWS)~\cite{aws}, Google Cloud Platform (GCP)~\cite{googlecloud}, and Microsoft Azure~\cite{azure}. They provide maximum convenience for users to easily deploy what they need to develop a service. In this context, they are not a platform to support the machine learning tasks, but they provide services that can easily apply or develop machine learning using their own powerful cloud resources and frameworks. However, these are not implemented solely for machine learning, so they can not overcome the inherent limitations. Commonly, they are very dependent on the machine learning framework what they use, and they are too free to construct a system, too much to do, or too repressed to have low degrees of freedom. Because they were developed as a commercial service, in addition, the expense for use might cause financial issues for researchers.

\section{Key features}
\subsection{System Overview}
\begin{figure}[ht]
\centering
\includegraphics[width=0.5\textwidth]{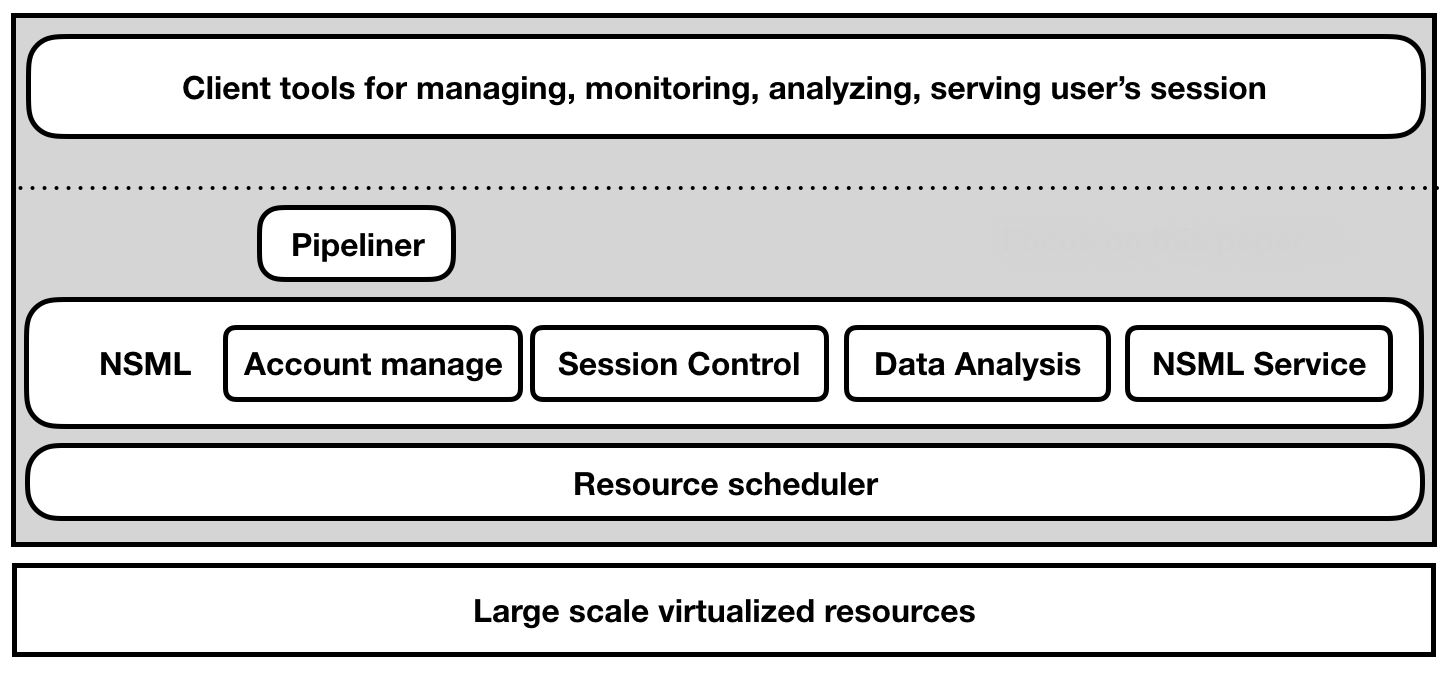}
\caption{High level overview of NSML}
\label{fig:overview}
\end{figure}

The overall illustration of NSML platform is presented in Figure~\ref{fig:overview}. NSML can be separated into two main modules for resource management and user interaction. The first part is presented at the bottom half of Figure~\ref{fig:overview} including \textit{Resource scheduler} and \textit{Large scale virtualized resources}. The resource management module in NSML should utilize a cluster with high computing resources, efficiently and flexibly. Once a computing node is registered in NSML ecosystem, its system resources will be regularly checked by NSML. If a user starts a machine learning job, NSML scheduler assigns the job to proper resources among the computing nodes which are registered.
The other part of NSML is user interface, shown in the top half of the figure from \textit{Client tool} to \textit{NSML}, via both command line and web-based which enhance the user experiences and conveniences. Users can control and manage a pipeline of their machine learning jobs, and obtain the models and experimental results through user interface. Additionally, serving APIs allows the users to design their own service using their models trained on NSML.

\subsection{Resources and Management}
Resource management is an important factor in developing these kinds of platform. Most of recent platforms are different from the ones that were managed on a resource dependent on each machine. They are a tendency to aggregate the resources held together and virtualize the resources. Also, the platforms distribute and manage the virtualized resources. This policy allows platform providers to operate their platform extensively with limited resources. For example, when a user makes a large number of requests that can not be handled by a single machine, this problem is solved by using virtualized computing resources from the platform. Another advantage is scalability. They can add resources while the platform is running.

One of the most popular virtualization methods for efficient resource management is Linux Containers~\cite{helsley2009lxc}. Prior to the appearance of the container technology, the environment was a configuration problem, and only a few kinds of app or framework was available in one Linux kernel. However, it is easier to manage a frameworks or app that uses resources by suggesting an idea that a container shares resources with a single kernel. Furthermore, the technology has been attract attention with the coming of a docker\footnote{https://www.docker.com/} that makes it easier to use.

Among a lot of management tools to apply the container, one of the most famous tool is Kubernetes~\cite{Hightower:2017:KUR:3175917}. Kubernetes makes it easy to manage complex container deployment tasks, but unfortunately it is not so easy to use that. The use of kubernetes for simple task is complex, and we have to consider too many situations. NSML focus on how to efficiently manage resources for machine learning rather than the complex production flow of other platforms, and the management policy for presented platform describe next subsections.

\subsubsection{Locality-Aware scheduling and residual resource defragmentation}
By the tendency of the modern machine learning algorithms using a massive amount of training data and computing resources, a scheduler should consider the data locality and defragmentation of the resource for maximizing the utilization. Unlike a dedicated server configurations, preparing dataset and environment often becomes performance bottleneck in shared-cluster setup. Moreover, it gets worse as the number of computing nodes grow.

In order to address this problem, we implemented our own scheduler--\textit{NSML scheduler} that considers the locality in the residence data and container images to mitigate overhead on data loading and environment setup. In addition, NSML scheduler manages residual resource defragmentation for important computing resources, especially GPUs. If a task asks for GPUs, then the scheduler sorts the available node list in ascending order by the number of available GPUs, and searches for a candidate from the first. Consequently, a node which has the largest number of GPUs may remain until the others are allocated.

\subsubsection{Scheduler fault tolerance}
Since the requests of users converge on a NSML scheduler node, a centralized system may be vulnerable to a single point of failure (SPOF)~\cite{armbrust2010view}. To avoid this failure, NSML scheduler consists of a primary and a secondary node as a main and a backup node, respectively. Although this warm-standby backup scheduler may overuse the computing resources, it can guarantee robustness against the failure of the primary scheduler. 

\subsubsection{Session and Resource Monitoring}
We installed two different kind of monitors for each computing node. The first one is a resource monitor, which collects the usage of computing resources (\textit{i.e.} CPU cores and GPUs) and periodically records them into a database. NSML scheduler assigns computing nodes to a requested job based on these information and requirements from the job. As shown in Figure~\ref{fig:gpu_utils}, the information collected by resource monitoring is also provided, so that users can check their models' efficiency by utilization of the assigned computing resource.

The other one is a session monitor for checking NSML sessions. It is also installed and running on computing nodes. When a user's session has been facing with an unexpected situation, the session monitor alarms the failure and the status of the session to the reporting chain. Then, our warning and control system informs the users of what is happened at their session via emails immediately, so that the users can handle the troubles. Additionally, the session monitor can keep track of status and progress of their session.

\subsection{Datasets and Collaboration}
\begin{figure}[tp]
\centering
\includegraphics[width=0.48\textwidth]{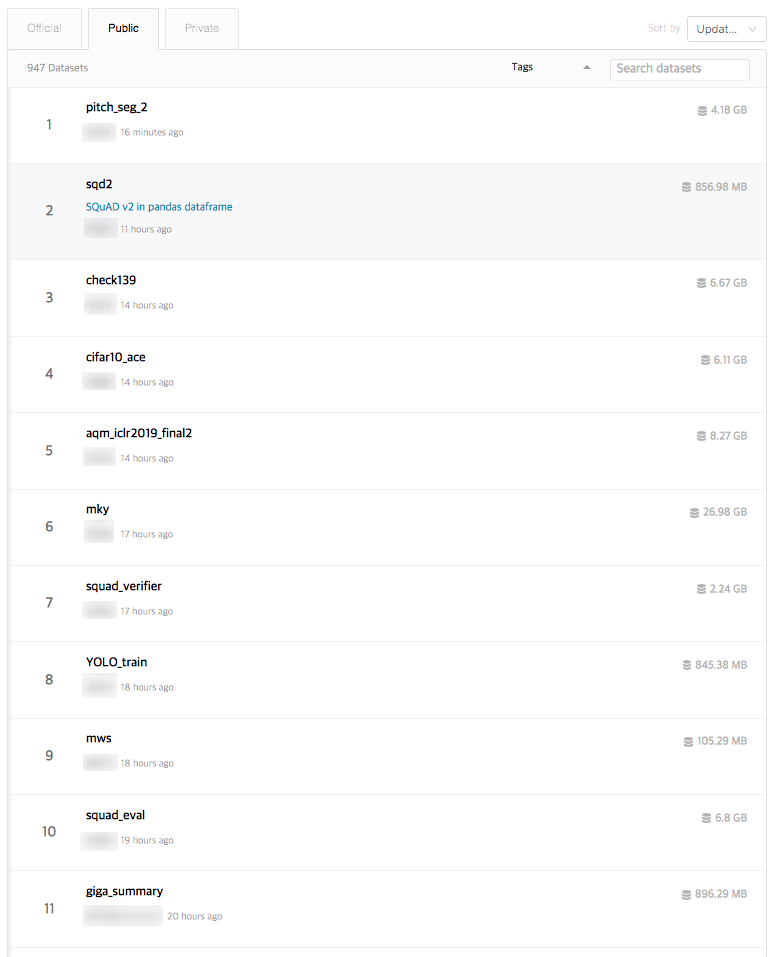}
\caption{Dataset view in NSML web application}
\label{fig:web_dataset}
\end{figure}
NSML provides functions for managing datasets and users. The datasets can be pushed to the public repository managed by NSML. If a user requests a machine learning job with a specified dataset not existing on the allocated node yet, the dataset will be copied into the node on demand during building an environment. After the dataset is cached in the node, a job which requires that dataset can start immediately without copying. If the owner sets the dataset to be public, it can be accessed by all users without any limitation. Consequently, if a user launches a job with the identical program code and dataset, the results can be easily reproduced.

Additionally, users can organize teams for collaboration. They can share their sessions, results, models, and program codes with team members. The users can set the private dataset within the team if needed so that only team members with a permission are able to access the dataset. Figure~\ref{fig:web_dataset} shows an example view of dataset list, and a user can check the meta-data of the registered dataset like the size and the latest access time.

\begin{table*}[ht]
\caption{User Client Tool}
{\small
\hfill{}
\begin{tabular}{|c|c|c|c|c|c|c|c|}
\hline
\textbf{Purpose}& \multicolumn{1}{|c|}{\textbf{Commands}}\\
\hline
Account Manage&credit, login, logout\\
\hline
Session Control&backup, command, diff, download, fork, getid, logs, ps, resume, rm, run, stop \\
\hline
Data Analysis&eventlen, events, exec, memo, model, plot, pull, sh, submit\\
\hline
NSML Service&automl, dataset, gpumornitor, gpustat, infer, status\\
\hline
\end{tabular}}
\hfill{}
\label{tb:clitool}
\end{table*}

\subsection{User Interface}
\subsubsection{Command line interface}
\begin{figure}
\begin{subfigure}{\linewidth}
  \centering
  \includegraphics[width=1.0\linewidth]{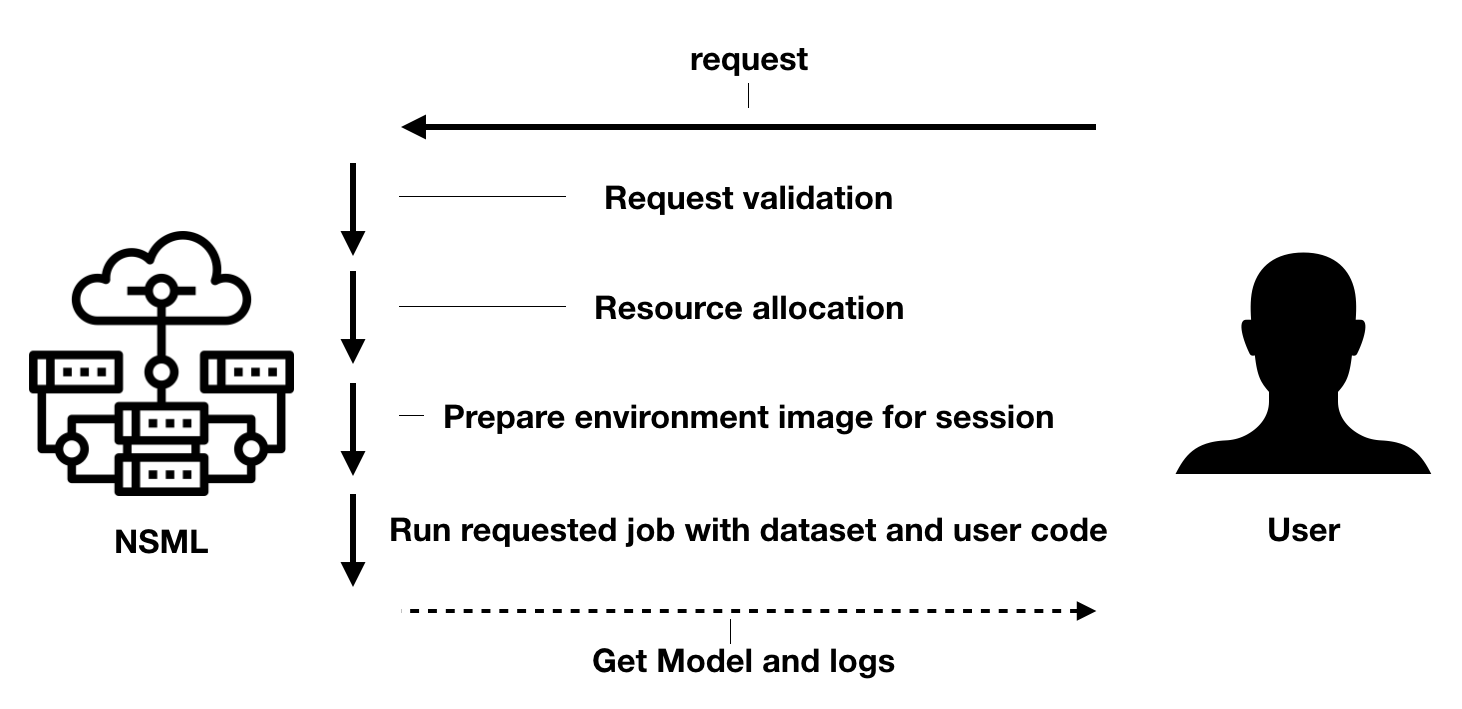}\hfill
  \caption{}
  \label{fig:workflow_a}
\end{subfigure}\par\medskip
\begin{subfigure}{\linewidth}
  \centering
  \includegraphics[width=1.0\linewidth]{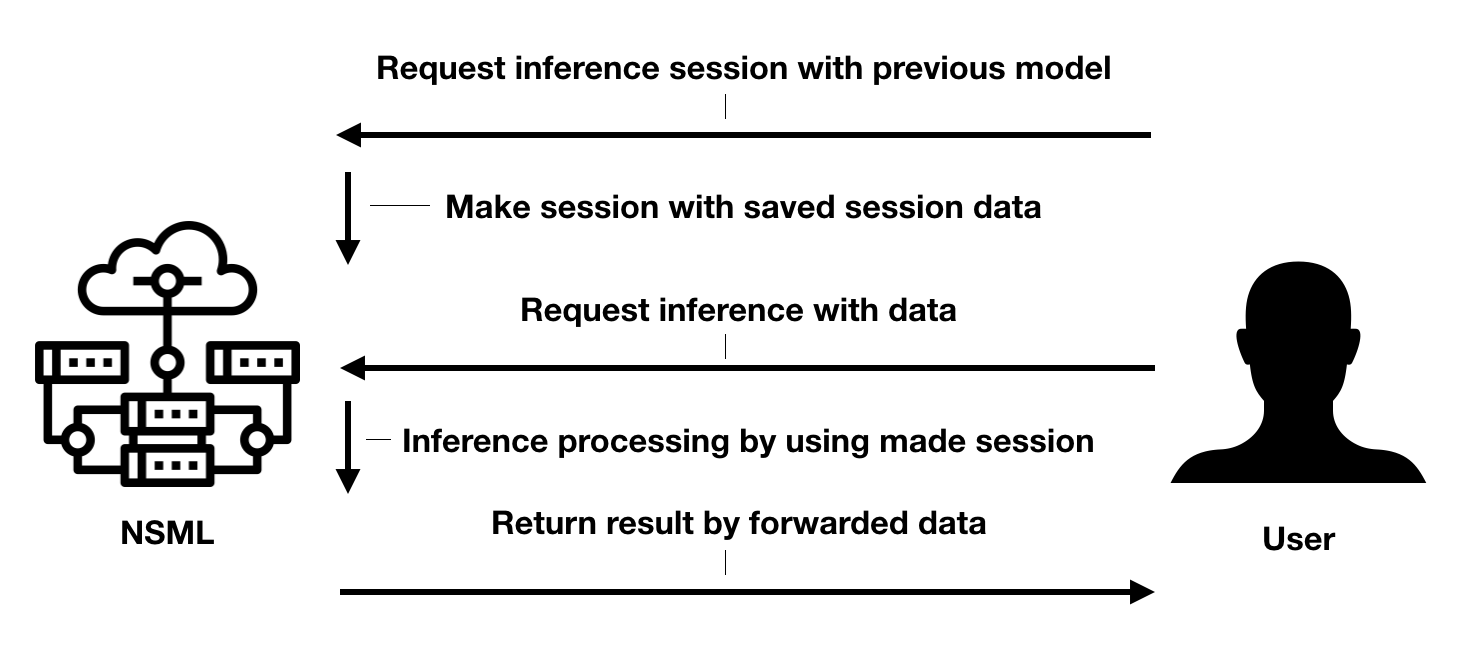}\hfill
  \caption{}
  \label{fig:workflow_b}
\end{subfigure}\par\medskip
\caption{Example of client tool command executing sequence}
\label{fig:workflow}
\end{figure}
Basically, NSML supports a series of commands through the command line interface (CLI). CLI does not require a specific operating system or environment, thus it can improve user experiences with platform independence. Figure~\ref{fig:workflow_a} and~\ref{fig:workflow_b} show the processes of \textit{run} and \textit{infer} commands, respectively. Existing code can be launched on NSML with a few additional lines. As shown in Table~\ref{tb:clitool}, all NSML commands belong to the following four categories. Some commands are only available to a system administrator.

\begin{description}
\item[Account Management]: 
While NSML provides only authorization like \textbf{login} for normal users, the administrators can manage the users' properties including the permission level and the credit through this category. The credit is used to regulate the monopolized usage of the cluster by a specific user. It is consumed when the user runs sessions according to the credit policy. If the credit is exhausted, the existing sessions may be safely stopped and the user cannot launch any more sessions.
\item [Session Control]:
The user can operate own code in a unit of a session with the resources in NSML. The request for system resources is estimated by the NSML scheduler, and can be rejected when the available resources are insufficient to meet the requirements. The session has saved all the information a user used including codes, logs, and meta-information, so that the user can reproduce or revise the owned sessions from the previous setting (\textit{e.g.} hyperparameter).
\item [Data Analysis]: 
These commands are used to collect the progress and results from a session. Using the collected data, the user can visualize learning curves, obtain trained models, and compare the results to other sessions. We described the details on visualization in Section~\ref{txt:vis}.
\item [NSML Service]:
NSML supports a number of additional services, for instance, allowing users to easily machine-learning work with remote resources. This is discussed in detail later.
\end{description}

\subsubsection{Web Interface and Visualization}\label{txt:vis}

\begin{figure*}[ht]
\centering
\includegraphics[width=0.95\textwidth]{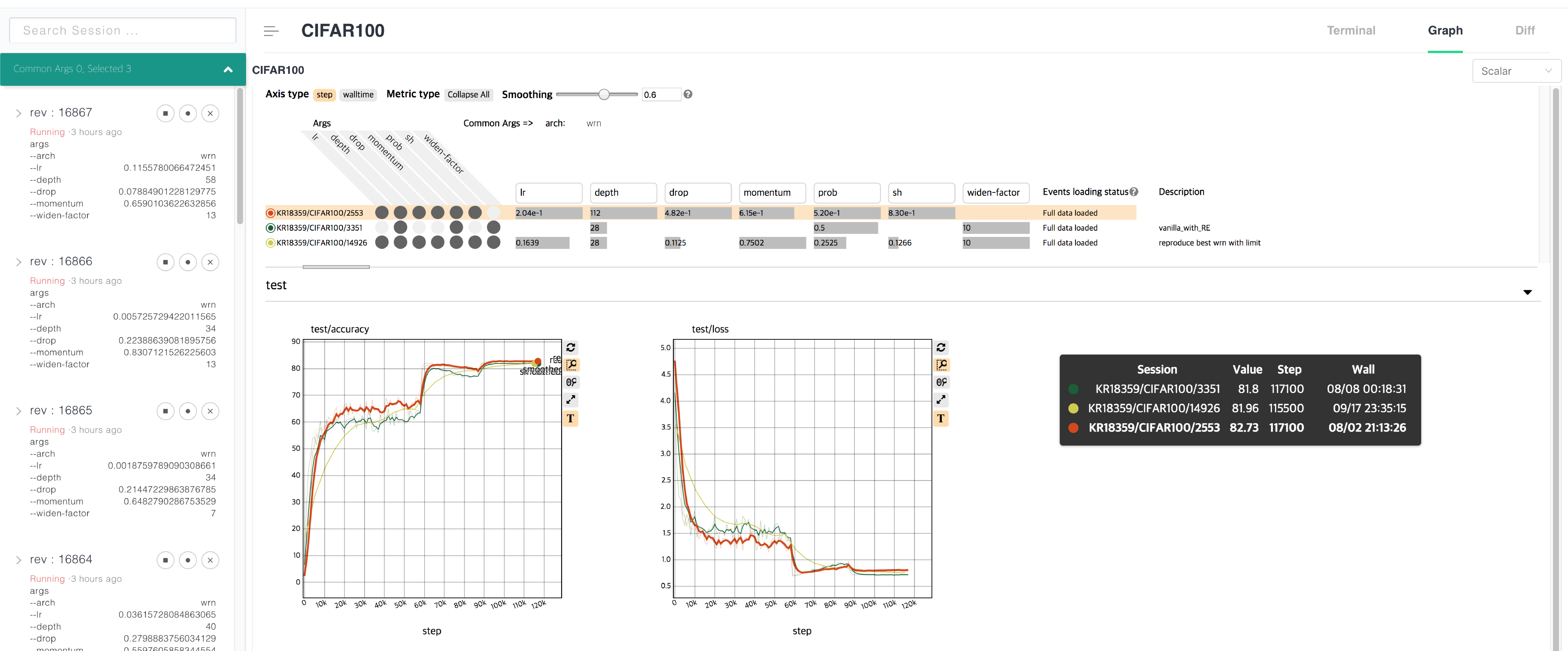}
\caption{Example of NSML analysis view: three sessions with different hyperparameters shown on NSML scalar plot. Top: The comparison part of each session. Bottom: Each plot of event data which users reported. The orange colored session is highlighted by the mouse on the `test/accuracy' plot.} 
\label{fig:web_analysis}
\end{figure*}

To improve user experience and system usability, we implemented web application of NSML.
The web app helps users to more intuitively and easily perform the operations for sessions: stopping, forking, logging, and deleting,  which could also be performed in the NSML CLI environment. Users can compare and analyze multiple sessions simultaneously through NSML web app. We also introduce powerful tools for visualizing the progress of users' sessions. Users can represent the results of their sessions as graphs by Visdom\footnote{\url{https://github.com/facebookresearch/visdom}} and TensorBoard~\cite{girija2016tensorflow}. Especially, our scalar plot has some advantages in comparing multiple sessions with different setup, as shown in Figure~\ref{fig:web_analysis}. 

As represented at the top-right side of Figure~\ref{fig:web_analysis}, the comparison panel is provided for comparing the arguments used in multiple sessions easily. The visualization window is composed of `common arguments' and `exclusive arguments' part to represent the arguments of each session. For the `exclusive arguments', a matrix-based diagram represents the different arguments in a scalable way. Through the diagram, users can easily recognize the overview of the differences between each session's argument such as whether the arguments exist in the session or not. Additionally, users can select and analyze the arguments in which they are interested, from the right side of the diagram. Also, by highlighting in each line graph when mouse interacted on the row of the visualization, the user recognizes which session is indicated currently.

Meanwhile, if multiple users work on the platform, web interface also provides administration features such as managing GPU usage of the total platform and checking resource usage of the specific session. Web interface also visualizes ranks among test results of the sessions so that users can easily compare their session performances and arguments. This feature will be useful for competitions.
 
\subsubsection{Serving API}
NSML platform provides serving APIs for the trained models as well as developing the models. The user can launch a session for serving the model created on the NSML using the command included in the client tool, and also test the inference modules to confirm the action of the submitted models. 

Furthermore, NSML is capable of serving through a RESTful API in order to improve accessibility and expand the application layers of NSML with web-based services. For example, if a user wants to serve their model directly on the Web, the user trains the model on the NSML platform, and simply submits their own inference procedure to the platform. At the service start time, the user starts the session with the submitted procedure for end-users to access the application that uses the trained models.

\subsection{Hyperparameter tuning}
Optimizing hyperparameter is one of the most time-consuming step of building a machine learning model. As the number of hyperparameter grows, the size of search space grows exponentially, and thus finding the best hyperparameters gets more difficult as well. As other ML platforms, NSML also supports the parallelized hyperparameter tuning. It is not only constrained to grid or random search, but also possible to apply many state-of-the art tuning algorithms such as population based training~\cite{jaderberg2017population}.

\subsection{Implementations}
NSML consists of three core modules; high computing resources, server-side NSML, and a scheduler. We chose Docker~\cite{docker} to set an environment for machine learning instances. Docker container is known as stable, light, and low-latent virtualization layer comparing to virtual machine\cite{felter2015updated} or virtual environment for python~\cite{reitz2016hitchhiker}. Furthermore, since Docker supports a network connection to communicate containers in a machine from another machine, we can control the sessions over the network. Consequently, many features will be available for session management such as stopping or removing a session with Docker. Containerized ML environment can guarantee stability as well as reproducibility, since it is able to manage the crucial system resources like memory or processors directly. 

Modern machine learning tasks require a lot of the system resources, and it may damage or even drive the host system into a panic by using excessive usage of the resource. In NSML, we used a virtualization layer to manage the resources assigned to each session, independently and exclusively. If a certain session claims too much memory over exceeding the allocated amount, it will be killed immediately by NSML without any damage on the host machine.

Even without the layer, server-side NSML contains many external components like database, messaging queue, and shared file system. Since many existing libraries used in NSML support python bindings, including machine learning applications, NSML was written in python primarily. The way of current implementation contributes sufficient flexibility to NSML flexible by providing more alternatives for the components. For instance, a system administrator may want to use any other kinds of DB schema such as SQL and NoSQL for NSML, and it is replaceable depending on how NSML manager will set their infrastructure because almost all frameworks and systems support python API. In addition, some part of the command-line interface layer was implemented in Go~\cite{donovan2015go} due to security and low latency.

We decided to design and implement our own scheduler and resource manager instead of using 3rd party libraries like docker swarm, since NSML scheduler asks a fine-grained management for the host machine's resource including GPUs. The scheduler part was developed in python for readily integrating to the entire system, moreover, it is running on the virtualization layer using docker for easy deployment and maintenance.

\begin{table*}[t]
\caption{Training Example Summary}
\centering
\begin{tabular}{lrrrrr}
\hline\hline
Dataset & Epochs & Batch size & Initial LR & Best Accuracy (\%) & Previous work (\%) \\
\hline\hline
MNIST  & 10 & 128 & 0.5 & 99.2 & 99.1~\cite{mnist_paper} \\
\hline
CIFAR-100 & 300 & 128 & 0.1 & 76.44 & 76.27~\cite{zhong2017random} \\
\hline
ImageNet & 120 & 2,048 & 0.8 & 74.8 & 75.3~\cite{resnet} \\
\hline\hline
\end{tabular}
\label{tab:example_summary}
\end{table*}

\begin{figure}[tp]
\centering
\includegraphics[width=0.48\textwidth]{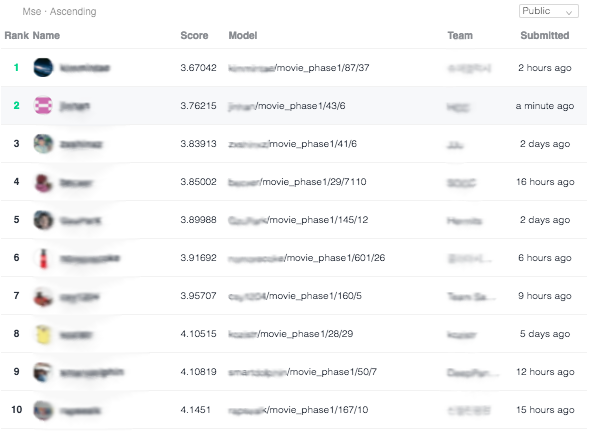}
\caption{Example of NSML leaderboard}
\label{fig:leaderboard}
\end{figure}

\begin{figure}[tp]
\centering
\includegraphics[width=0.48\textwidth]{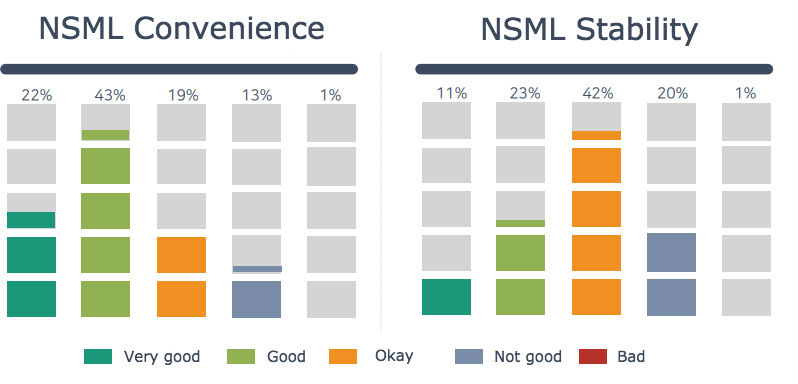}
\caption{Feedback from competition participants}
\label{fig:feedback}
\end{figure}

\begin{table*}[ht]
\caption{User statistics in NLP competition}
\centering
\begin{tabular}{l rrrrr}
\hline\hline
& Questions stage 1 & Questions stage 2 & Movie stage 1 & Movie stage 2 \\[1ex]
\hline\hline
\# of users  & 93 & 89 & 55 & 58  \\[1ex]
\hline
\# of models & 3,907 & 3,658 & 5,044 & 5,267 \\[1ex]
\hline
Average \# of models per user & 42.01 & 41.10 & 91.71 & 90.81 \\[1ex]
\hline
Max \# of models per user & 329 & 383 & 1,103 & 732 \\[1ex]
\hline
\# of users less than 5 models & 24 & 27 & 14 & 13 \\[1ex]
\hline\hline
\end{tabular}
\label{tab:nlp_hackathon}
\end{table*}

\begin{figure*}[ht]
\centering
\includegraphics[width=0.96\textwidth]{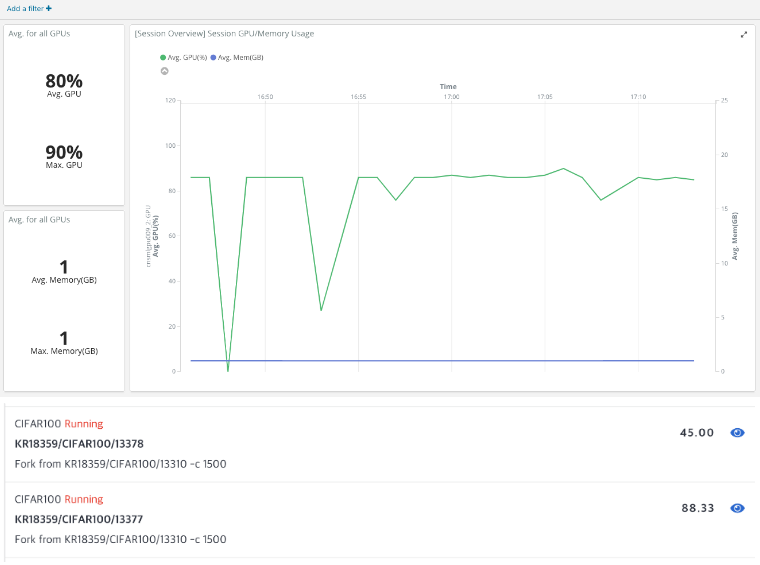}
\caption{Monitoring on GPUs for each session. Top: History of GPU utilization of one session. Bottom: Dashboard with average GPU utilization of each session}
\label{fig:kibana}
\end{figure*}

\begin{table*}[ht]
\caption{User statistics in Angle prediction and Keyboard correction competitions}
\centering
\begin{tabular}{l rr}
\hline\hline
& Angle Prediction & Keyboard Correction \\[1ex]
\hline
Average \# of models per user & 78.87 & 93.18 \\[1ex]
\hline
Max \# of models per user & 546 & 1,075  \\[1ex]
\hline
Ratio of users less than 5 models & 53.3\% & 50.8\% \\[1ex]
\hline
Average \# of submissions per user& 63.57 & 84.94 \\[1ex]
\hline
Max \# of submissions per user& 553 & 2,008 \\[1ex]
\hline
Ratio of users less than 5 submissions & 56.7\% & 56.6\% \\[1ex]
\hline\hline
\end{tabular}
\label{tab:other_hackathons}
\end{table*}

\begin{figure}[ht]
\includegraphics[width=0.48\textwidth,trim={5.5cm 2.2cm 5.5cm 2.2cm},clip]{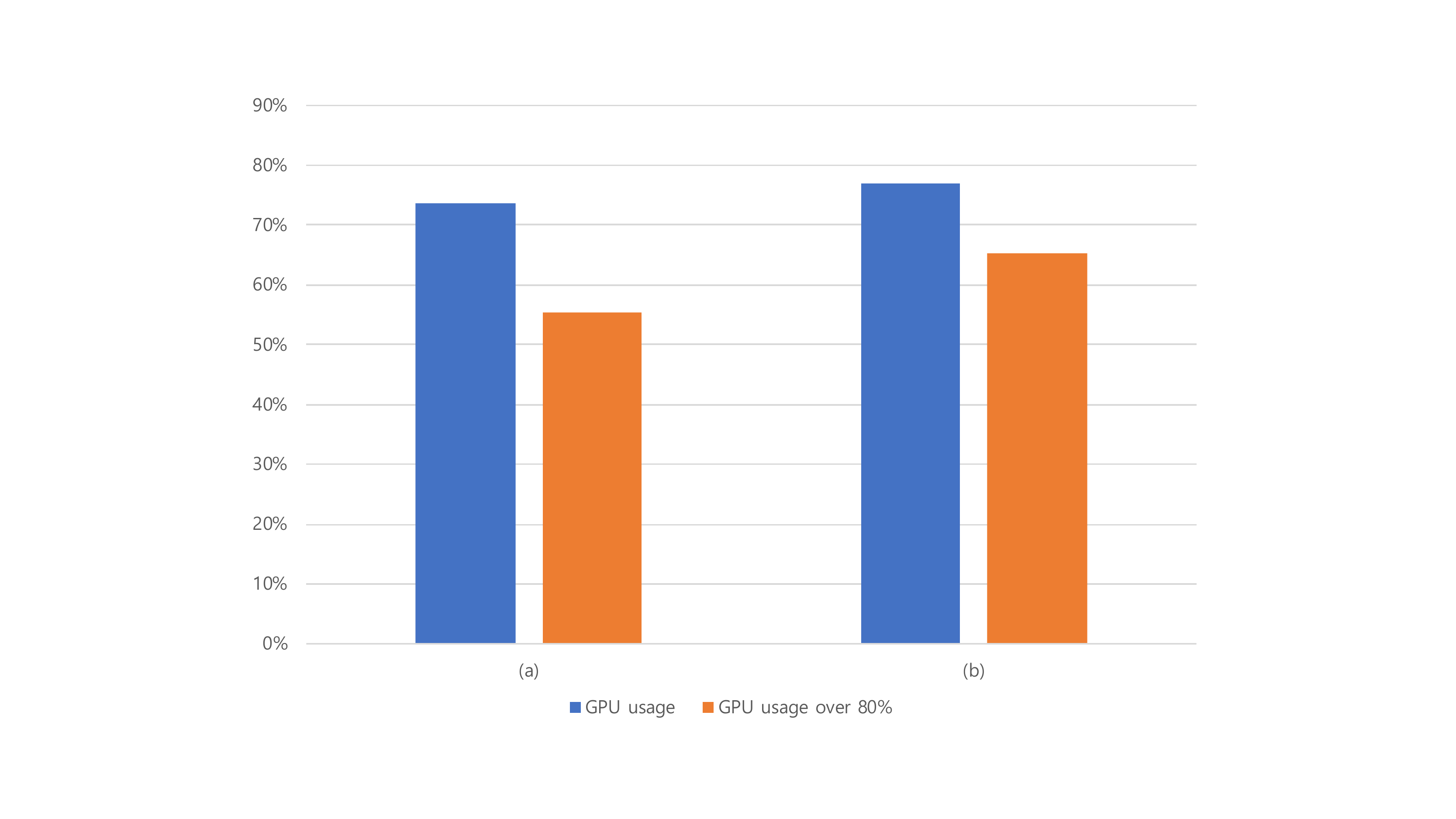}
\centering
\caption{Monitoring ratio of running GPUs. Blue bars present ratio of all running GPUs, and orange bars show ratio of GPUs with over 80\% of utilization. (a) Before supporting visualization of GPU utilization, (b) After supporting GPU utilization}
\label{fig:gpu_utils}
\end{figure}

\section{Experiments and Case Studies}
\subsection{Experimental Setup}
Our proposed NSML has been running on our computing cluster, which includes $n$ computing nodes with hundreds of GPUs. Each computing node has up to 256GB of system memories and 8 GPUs. To demonstrate the reproducibility of NSML, we conducted the training of the existing models for image classification datasets; MNIST~\cite{mnist}, CIFAR-100~\cite{cifar}, and ImageNet~\cite{imagenet2012}. All the programming codes are written in python using PyTorch~\cite{pytorch} and Torchvision~\cite{torchvision} libraries. As a environmental setup, we allocated one GPU for both MNIST and CIFAR-100 each, and eight GPUs for a model on ImageNet.

For MNIST example, we designed a simple convolutional neural network which consisted of two convolutional layers and two inner product layer. We achieved 99.22 percent of accuracy for the validation set of MNIST example, and it is acceptable level according to the previous work~\cite{mnist_paper}. We also trained ResNet-110 and ResNet-50 models for CIFAR-100 and ImageNet datasets using the previous work~\cite{zhong2017random,resnet}, respectively. The result explains that we reproduce the previous work using NSML successfully. The summary of the result is shown in Table~\ref{tab:example_summary}.

\subsection{ML Competitions via NSML}
We have held three machine learning competitions on our proposed NSML. Each task is related to the followings; 1) General natural language tasks, 2) Predicting tilted angle of given document images, and 3) Correcting keyboard typo on mobile devices, respectively. Except the first competition, the best models in last two competitions perform better than the baseline which were on real customer services. After the competitions were finished, the best models have been applied to enhancing the services. The entire pipeline of competitions are; preprocessing for data to result submission and test, all the processes were running on NSML, and we also supported real-time leaderboard on our website so that all participants could immediately check their ranking.

Figure~\ref{fig:leaderboard} presents an example of NSML leaderboard from the first competition. The figure shows the list of user ID, dataset, ranking, score, and name of evaluation metric and order. In addition, it is able to display submission history for each user to check trend of model performance.

\subsubsection{Natural Language Tasks}
We suggested two problems for the natural language processing competition; 1) Predicting preference score of movie from user's review and 2) Recognizing if two sentences are asking the same questions or not. The former one is a kind of regression task, thus mean squared error~\cite{LehmCase98} was chosen as an evaluation metric. For the latter one, we used accuracy as an evaluation metric since the latter task is binary classification.

This competition was composed of three rounds. First two rounds had been scheduled for two weeks via online respectively, with different size of datasets. The size of dataset used in the second round was twice as much as in the first round. The final round was held for two full days in offline with a new, but comparatively small datasets. The initial participants were a hundred teams selected from more than 1,300 users and 700 teams. The end of each round, only participants with higher score were survived.

During the competition, we have collected feedback and interview from our participants. Figure~\ref{fig:feedback} shows summary of NSML-related feedback from the user. The majority of participants have been satisfied with user interface of NSML. They were satisfied by NSML's usability to develop their models and visualization through web interface. However, they were not that satisfied on system robustness. According to the user interview, the most of dissatisfaction was caused by lack of resources, especially GPUs, and the other was because of system failure in networking or database.

Table~\ref{tab:nlp_hackathon} shows statistics of participants behavior. As shown in the table, most of users succeeded to run at least one session in the first phase. Consequently, we can conjecture that NSML does not have high entry barrier for majority of users. Only small ratio of participants launched less than 5 models, then, NSML proved its accessibility and user convenience for developing machine learning models.

\subsubsection{Object-Rotating Angle Prediction}
The objective of the second competition was to predict the angle from the given tilted images. Accuracy has been used for the evaluation metric with 1 degree of error range. The competition was held as a single-game without multi-rounds for a month. All participants were selected from our organization internally, without considering their background. Only few of the participants had a knowledge of machine learning or computer vision, and even, some of them were not educated in the regular course of computer science. The first column of Table~\ref{tab:other_hackathons} represents behavior of Angle prediction competition. Only slightly above a half of participants used our framework very rarely (less than 5 training and submissions), which means other half of them used NSML frequently for training.

\subsubsection{Keyboard-Typo Correction}
For the latest competition, the examination problem was to correct an invalid keystroke from the given sequence of keyboard inputs on a mobile device. The model which participants should design decides whether the given keyboard input is correct or not, and if not, corrects it. F1 measure~\cite{Chinchor:1992:MEM:1072064.1072067} has been used for the evaluation metric. This competition also consisted of only a single round, and was held for two and a half months. Similar to the competition for angle prediction, all participants were selected from the our company internally without considering their background.

The second column of Table~\ref{tab:other_hackathons} represents information for Keyboard correction competition. The numbers of the competition was analogous to the previous one, and at least over a half of participants still continued to develop their model using our platform after the middle of competition. It implies that NSML shows its convenience for improving the existing ML works as well as the lower entry barrier for starter.

\section{Discussion}
\subsection{Resource Utilization}
Currently, hundreds of users perform their ML work on NSML which is running on a cluster with massive computing resources. In the last several months, utilization of our resources, the ratio of running GPUs to the total GPUs is over $70\%$ on average, and especially, is achieved full utilization ($\approx 100\%$) at the peak. Even though the number of entire GPUs keeps increasing in the last a few months, average of GPU utilization along time has not been decreased. Consequently, usability of NSML is remarkable, especially on dealing with many experiments demanding massive computing resources.

Meanwhile, even though the users want to optimize their code in terms of resource efficiency, they cannot achieve without knowing the GPU usage of their model. For easy profiling and monitoring, we also introduced the monitoring tools which collects and visualizes the utilization of a single GPU automatically via Kibana~\cite{kibana}. Figure~\ref{fig:kibana} shows both history of utilization and occupied memory of one GPU and dashboard for average GPU utilization on the latest NSML sessions. The bottom side is shown on the front web page, so the users are informed average GPU usage of their sessions. The top image would be displayed if the users need information in detail, and with the image, they would be able to check which line of code harms the resource efficiency.

The utilization has had a tendency to increase consistently ever since the monitoring feature was active. With the feature, the users can figure out efficiency of code and try to improve it themselves. And they actually increased the number of GPUs with over 80\% of GPU utilization. As it is seen in Figure~\ref{fig:gpu_utils}, percentage of running GPUs has increased less than 5\%, while percentage of GPUs with over 80\% utilization has increased about 10\% after the visualization feature. We surmised that the users may improve their strategy for GPU usage referring to the collected information.

\subsection{Limitations and Future Work}
Although we have succeeded to improve resource efficiency and usability for ML research through NSML, a few limitations are still left. Above all, one of the most important concerns is related to managing datasets efficiently. Although NSML supports several file systems including shared file system over networking, however, the absence of version control for datasets makes it difficult to even patch a minor revision on the existing one. Moreover, as multi-task learning and transfer learning are expected to become popular, the demand for using multiple datasets grows gradually. However, a single task using the multiple dataset is not yet supported due to the limitation from the current design of NSML's session representation. We are considering these features as a top priority, and make it available in the near future. 

The other important issue is an advanced visualization features. Though the existing visualization tools provide abundant illustrations for the scalar plotting, NSML is still unable to express in a various media including visualization for the complicated concepts, audio/video output, embedding layer, feature maps, and any other complex representations in an easy way. 

In addition, we are implementing the distributed learning feature on NSML, which will provide the parallelism over multiple nodes. With this feature, a user can deploy a gigantic dataset or model which are difficult to load into a single machine. We expect it to enhance the overall system utilization and relief the lack of resources caused by fragmented resource allocation.

\section{Conclusion}
As the application of machine learning is tightly coupled with industry and commercial services, the efficient developing environments and its infrastructure become more important for large scale clusters which many companies and organizations already owned. In this paper, we proposed NSML as a novel platform for machine learning as a service (MLaaS), and demonstrated its potential with some examples and use cases. NSML also provides comfortable tools and a serverless foundation to users as well as collaborative environments. With several competitions on NSML, especially, we convinced that NSML can be a key player for commercialization of machine learning models. 

%% file: nsml_main.bbl
\begin{thebibliography}{32}
\providecommand{\natexlab}[1]{#1}
\providecommand{\url}[1]{\texttt{#1}}
\expandafter\ifx\csname urlstyle\endcsname\relax
  \providecommand{\doi}[1]{doi: #1}\else
  \providecommand{\doi}{doi: \begingroup \urlstyle{rm}\Url}\fi

\bibitem[Abadi et~al.(2016)Abadi, Barham, Chen, Chen, Davis, Dean, Devin,
  Ghemawat, Irving, Isard, Kudlur, Levenberg, Monga, Moore, Murray, Steiner,
  Tucker, Vasudevan, Warden, Wicke, Yu, and Zheng]{tensorflow}
Abadi, M., Barham, P., Chen, J., Chen, Z., Davis, A., Dean, J., Devin, M.,
  Ghemawat, S., Irving, G., Isard, M., Kudlur, M., Levenberg, J., Monga, R.,
  Moore, S., Murray, D.~G., Steiner, B., Tucker, P., Vasudevan, V., Warden, P.,
  Wicke, M., Yu, Y., and Zheng, X.
\newblock Tensorflow: A system for large-scale machine learning.
\newblock In \emph{Proceedings of the 12th USENIX Conference on Operating
  Systems Design and Implementation}, OSDI'16, pp.\  265--283, Berkeley, CA,
  USA, 2016. USENIX Association.
\newblock ISBN 978-1-931971-33-1.
\newblock URL \url{http://dl.acm.org/citation.cfm?id=3026877.3026899}.

\bibitem[Armbrust et~al.(2010)Armbrust, Fox, Griffith, Joseph, Katz, Konwinski,
  Lee, Patterson, Rabkin, Stoica, et~al.]{armbrust2010view}
Armbrust, M., Fox, A., Griffith, R., Joseph, A.~D., Katz, R., Konwinski, A.,
  Lee, G., Patterson, D., Rabkin, A., Stoica, I., et~al.
\newblock A view of cloud computing.
\newblock \emph{Communications of the ACM}, 53\penalty0 (4):\penalty0 50--58,
  2010.

\bibitem[Baylor et~al.(2017)Baylor, Breck, Cheng, Fiedel, Foo, Haque, Haykal,
  Ispir, Jain, Koc, et~al.]{baylor2017tfx}
Baylor, D., Breck, E., Cheng, H.-T., Fiedel, N., Foo, C.~Y., Haque, Z., Haykal,
  S., Ispir, M., Jain, V., Koc, L., et~al.
\newblock Tfx: A tensorflow-based production-scale machine learning platform.
\newblock In \emph{Proceedings of the 23rd ACM SIGKDD International Conference
  on Knowledge Discovery and Data Mining}, pp.\  1387--1395. ACM, 2017.

\bibitem[Boettiger(2015)]{Boettiger:2015:IDR:2723872.2723882}
Boettiger, C.
\newblock An introduction to docker for reproducible research.
\newblock \emph{SIGOPS Oper. Syst. Rev.}, 49\penalty0 (1):\penalty0 71--79,
  January 2015.
\newblock ISSN 0163-5980.
\newblock \doi{10.1145/2723872.2723882}.
\newblock URL \url{http://doi.acm.org/10.1145/2723872.2723882}.

\bibitem[Chhajed(2015)]{kibana}
Chhajed, S.
\newblock \emph{Learning ELK Stack}.
\newblock Packt Publishing Ltd, 2015.

\bibitem[Chinchor(1992)]{Chinchor:1992:MEM:1072064.1072067}
Chinchor, N.
\newblock Muc-4 evaluation metrics.
\newblock In \emph{Proceedings of the 4th Conference on Message Understanding},
  MUC4 '92, pp.\  22--29, Stroudsburg, PA, USA, 1992. Association for
  Computational Linguistics.
\newblock ISBN 1-55860-273-9.
\newblock \doi{10.3115/1072064.1072067}.
\newblock URL \url{https://doi.org/10.3115/1072064.1072067}.

\bibitem[Copeland et~al.(2015)Copeland, Soh, Puca, Manning, and Gollob]{azure}
Copeland, M., Soh, J., Puca, A., Manning, M., and Gollob, D.
\newblock \emph{Microsoft Azure: Planning, Deploying, and Managing Your Data
  Center in the Cloud}.
\newblock Apress, Berkely, CA, USA, 1st edition, 2015.
\newblock ISBN 1484210441, 9781484210444.

\bibitem[Diginmotion(2011)]{aws}
Diginmotion.
\newblock Amazon web services.
\newblock 2011.

\bibitem[Donovan \& Kernighan(2015)Donovan and Kernighan]{donovan2015go}
Donovan, A.~A. and Kernighan, B.~W.
\newblock \emph{The Go programming language}.
\newblock Addison-Wesley Professional, 2015.

\bibitem[Felter et~al.(2015)Felter, Ferreira, Rajamony, and
  Rubio]{felter2015updated}
Felter, W., Ferreira, A., Rajamony, R., and Rubio, J.
\newblock An updated performance comparison of virtual machines and linux
  containers.
\newblock In \emph{Performance Analysis of Systems and Software (ISPASS), 2015
  IEEE International Symposium On}, pp.\  171--172. IEEE, 2015.

\bibitem[Girija(2016)]{girija2016tensorflow}
Girija, S.~S.
\newblock Tensorflow: Large-scale machine learning on heterogeneous distributed
  systems.
\newblock 2016.

\bibitem[He et~al.(2016)He, Zhang, Ren, and Sun]{resnet}
He, K., Zhang, X., Ren, S., and Sun, J.
\newblock Deep residual learning for image recognition.
\newblock In \emph{2016 {IEEE} Conference on Computer Vision and Pattern
  Recognition, {CVPR} 2016, Las Vegas, NV, USA, June 27-30, 2016}, pp.\
  770--778, 2016.
\newblock \doi{10.1109/CVPR.2016.90}.
\newblock URL \url{https://doi.org/10.1109/CVPR.2016.90}.

\bibitem[Helsley(2009)]{helsley2009lxc}
Helsley, M.
\newblock Lxc: Linux container tools.
\newblock \emph{IBM devloperWorks Technical Library}, 11, 2009.

\bibitem[Hightower et~al.(2017)Hightower, Burns, and
  Beda]{Hightower:2017:KUR:3175917}
Hightower, K., Burns, B., and Beda, J.
\newblock \emph{Kubernetes: Up and Running Dive into the Future of
  Infrastructure}.
\newblock O'Reilly Media, Inc., 1st edition, 2017.
\newblock ISBN 1491935677, 9781491935675.

\bibitem[Jaderberg et~al.(2017)Jaderberg, Dalibard, Osindero, Czarnecki,
  Donahue, Razavi, Vinyals, Green, Dunning, Simonyan,
  et~al.]{jaderberg2017population}
Jaderberg, M., Dalibard, V., Osindero, S., Czarnecki, W.~M., Donahue, J.,
  Razavi, A., Vinyals, O., Green, T., Dunning, I., Simonyan, K., et~al.
\newblock Population based training of neural networks.
\newblock \emph{arXiv preprint arXiv:1711.09846}, 2017.

\bibitem[Krishnan \& Gonzalez(2015)Krishnan and Gonzalez]{googlecloud}
Krishnan, S.~T. and Gonzalez, J.~U.
\newblock \emph{Building Your Next Big Thing with Google Cloud Platform: A
  Guide for Developers and Enterprise Architects}.
\newblock Apress, Berkely, CA, USA, 1st edition, 2015.
\newblock ISBN 1484210050, 9781484210055.

\bibitem[Krizhevsky \& Hinton(2009)Krizhevsky and Hinton]{cifar}
Krizhevsky, A. and Hinton, G.
\newblock Learning multiple layers of features from tiny images.
\newblock Technical report, Citeseer, 2009.

\bibitem[Krizhevsky et~al.(2012)Krizhevsky, Sutskever, and Hinton]{deepimage}
Krizhevsky, A., Sutskever, I., and Hinton, G.~E.
\newblock Imagenet classification with deep convolutional neural networks.
\newblock In \emph{Advances in neural information processing systems}, pp.\
  1097--1105, 2012.

\bibitem[LeCun \& Cortes(2010)LeCun and Cortes]{mnist}
LeCun, Y. and Cortes, C.
\newblock {MNIST} handwritten digit database.
\newblock 2010.
\newblock URL \url{http://yann.lecun.com/exdb/mnist/}.

\bibitem[LeCun et~al.(1998)LeCun, Bottou, Bengio, and Haffner]{mnist_paper}
LeCun, Y., Bottou, L., Bengio, Y., and Haffner, P.
\newblock Gradient-based learning applied to document recognition.
\newblock \emph{Proceedings of the IEEE}, 86\penalty0 (11):\penalty0
  2278--2324, November 1998.

\bibitem[Lehmann \& Casella(1998)Lehmann and Casella]{LehmCase98}
Lehmann, E.~L. and Casella, G.
\newblock \emph{Theory of Point Estimation}.
\newblock Springer-Verlag, New York, NY, USA, second edition, 1998.

\bibitem[Liang et~al.(2018)Liang, Krishnan, Hoffman, and
  Jebara]{liang2018variational}
Liang, D., Krishnan, R.~G., Hoffman, M.~D., and Jebara, T.
\newblock Variational autoencoders for collaborative filtering.
\newblock \emph{arXiv preprint arXiv:1802.05814}, 2018.

\bibitem[Marcel \& Rodriguez(2010)Marcel and Rodriguez]{torchvision}
Marcel, S. and Rodriguez, Y.
\newblock Torchvision the machine-vision package of torch.
\newblock In \emph{Proceedings of the 18th ACM International Conference on
  Multimedia}, MM '10, pp.\  1485--1488, New York, NY, USA, 2010. ACM.
\newblock ISBN 978-1-60558-933-6.
\newblock \doi{10.1145/1873951.1874254}.
\newblock URL \url{http://doi.acm.org/10.1145/1873951.1874254}.

\bibitem[Merkel(2014)]{docker}
Merkel, D.
\newblock Docker: Lightweight linux containers for consistent development and
  deployment.
\newblock \emph{Linux J.}, 2014\penalty0 (239), March 2014.
\newblock ISSN 1075-3583.
\newblock URL \url{http://dl.acm.org/citation.cfm?id=2600239.2600241}.

\bibitem[Min et~al.(2017)Min, Lee, and Yoon]{deepbio}
Min, S., Lee, B., and Yoon, S.
\newblock Deep learning in bioinformatics.
\newblock \emph{Briefings in bioinformatics}, 18\penalty0 (5):\penalty0
  851--869, 2017.

\bibitem[Paszke et~al.(2017)Paszke, Gross, Chintala, Chanan, Yang, DeVito, Lin,
  Desmaison, Antiga, and Lerer]{pytorch}
Paszke, A., Gross, S., Chintala, S., Chanan, G., Yang, E., DeVito, Z., Lin, Z.,
  Desmaison, A., Antiga, L., and Lerer, A.
\newblock Automatic differentiation in pytorch.
\newblock In \emph{NIPS-W}, 2017.

\bibitem[Reitz \& Schlusser(2016)Reitz and Schlusser]{reitz2016hitchhiker}
Reitz, K. and Schlusser, T.
\newblock \emph{The Hitchhiker's Guide to Python: Best Practices for
  Development}.
\newblock " O'Reilly Media, Inc.", 2016.

\bibitem[Russakovsky et~al.(2015)Russakovsky, Deng, Su, Krause, Satheesh, Ma,
  Huang, Karpathy, Khosla, Bernstein, Berg, and Fei-Fei]{imagenet2012}
Russakovsky, O., Deng, J., Su, H., Krause, J., Satheesh, S., Ma, S., Huang, Z.,
  Karpathy, A., Khosla, A., Bernstein, M., Berg, A.~C., and Fei-Fei, L.
\newblock {ImageNet Large Scale Visual Recognition Challenge}.
\newblock \emph{International Journal of Computer Vision (IJCV)}, 115\penalty0
  (3):\penalty0 211--252, 2015.
\newblock \doi{10.1007/s11263-015-0816-y}.

\bibitem[Xiong et~al.(2018)Xiong, Wu, Alleva, Droppo, Huang, and
  Stolcke]{xiong2018microsoft}
Xiong, W., Wu, L., Alleva, F., Droppo, J., Huang, X., and Stolcke, A.
\newblock The microsoft 2017 conversational speech recognition system.
\newblock In \emph{2018 IEEE International Conference on Acoustics, Speech and
  Signal Processing (ICASSP)}, pp.\  5934--5938. IEEE, 2018.

\bibitem[Yamada et~al.(2018)Yamada, Iwamura, and Kise]{yamada2018shakedrop}
Yamada, Y., Iwamura, M., and Kise, K.
\newblock Shakedrop regularization.
\newblock \emph{arXiv preprint arXiv:1802.02375}, 2018.

\bibitem[Yu et~al.(2018)Yu, Dohan, Luong, Zhao, Chen, Norouzi, and
  Le]{yu2018qanet}
Yu, A.~W., Dohan, D., Luong, M.-T., Zhao, R., Chen, K., Norouzi, M., and Le,
  Q.~V.
\newblock Qanet: Combining local convolution with global self-attention for
  reading comprehension.
\newblock \emph{arXiv preprint arXiv:1804.09541}, 2018.

\bibitem[Zhong et~al.(2017)Zhong, Zheng, Kang, Li, and Yang]{zhong2017random}
Zhong, Z., Zheng, L., Kang, G., Li, S., and Yang, Y.
\newblock Random erasing data augmentation.
\newblock \emph{arXiv preprint arXiv:1708.04896}, 2017.

\end{thebibliography}
